# Combining half-metals and multiferroics into epitaxial heterostructures for spintronics

H. Béa

Unité Mixte de Physique CNRS-Thales, Route Départementale 128, 91767 Palaiseau, France

M. Bibes

Institut d'Electronique Fondamentale, Université Paris-Sud, 91405 Orsay, France

M. Sirena, G. Herranz, K. Bouzehouane, E. Jacquet

Unité Mixte de Physique CNRS-Thales, Route Départementale 128, 91767 Palaiseau, France

S. Fusil

Université d'Evry, Bâtiment des Sciences, rue du père Jarlan, 91025 Evry, France

P. Paruch, M. Dawber

DPMC, University of Geneva, 24 Quai E. Ansermet, 1211 Geneva 4, Switzerland

J.-P. Contour and A. Barthélémy

Unité Mixte de Physique CNRS-Thales, Route Départementale 128, 91767 Palaiseau, France

Abstract

We report on the growth of epitaxial bilayers of the $La_{2/3}Sr_{1/3}MnO_3$ (LSMO) half-metallic ferromagnet and the $BiFeO_3$ (BFO) multiferroic, on $SrTiO_3$(001) by pulsed laser deposition. The growth mode of both layers is two-dimensional, which results in unit-cell smooth surfaces. We show that both materials keep their properties inside the heterostructures, i.e. the LSMO layer (11 nm thick) is ferromagnetic with a Curie temperature of ~330K, while the BFO films shows ferroelectricity down to very low thicknesses (5 nm). Conductive-tip atomic force microscope mappings of BFO/LSMO bilayers for different BFO thicknesses reveal a high and homogeneous resistive state for the BFO film that can thus be used as a ferroelectric tunnel barrier in tunnel junctions based on a half-metal.

To a great extent, today's research in spintronics focuses on the development of materials and devices concepts [1]. In many cases, the former are requisites to the latter. Recently, several families of materials have been developed in this sense, a typical example being that of diluted magnetic semiconductors [2]. As spintronics effects rely on the spin polarization of the electrical current, materials with large, ideally total, spin-polarization have also been extensively investigated. With these so-called half-metals [3], a considerable increase in the tunnel magnetoresistance (TMR) of magnetic tunnel junctions (MTJs) [4] has indeed been achieved, at least at low temperatures [5].

Another emerging family of magnetic materials are multiferroics [6]. In these compounds, several ferroic orders (e.g. magnetic and electric) coexist, with some coupling between them (the magnetoelectric effect [7]). Most are ferroelectric and antiferromagnetic, a notable exception being $BiMnO_3$ that is ferromagnetic [8]. A propotypical multiferroic that has attracted a lot of attention lately is $BiFeO_3$ (BFO) [9,10]. It is a ferroelectric and weakly ferromagnetic rhombohedral perovskite with order temperatures far above room temperature ($T_C$=1043K [11], $T_N$=647K [12]). BFO thus crystallises in the same structure as several known half-metallic ferromagnets (such as $La_{2/3}Sr_{1/3}MnO_3$, $La_{2/3}Ca_{1/3}MnO_3$ or $Sr_2FeMoO_6$), which makes possible to combine it with these materials in multifunctional epitaxial heterostructures. Several promising types of devices can be imagined from this combination, as discussed for instance by Binek et al [13,14]. In particular, one can think of using very thin layers of $BiFeO_3$ as multiferroic tunnel barriers. If ferroelectric, these layers should have the same functionalities as those of recently developed [15] and modelled [16,17] ferroelectric tunnel barriers, combined with a magnetic ordering and a possible magnetoelectric coupling.

In this paper, we describe the growth and properties of bilayers of the $La_{2/3}Sr_{1/3}MnO_3$ (LSMO) half-metal combined with BFO and epitaxially grown onto $SrTiO_3$(001) substrates by pulsed laser deposition. We study the morphological, structural, electrical and magnetic properties of BFO/LSMO bilayers. We show that the magnetic properties of LSMO are preserved and that the BFO layers are insulating and ferroelectric, down to thicknesses of $t_{BFO}$=5 nm. BFO ultrathin layers thus fulfil some important criteria for being used as ferroelectric tunnel barriers.

The samples used in this study have been grown by pulsed laser deposition using a Nd:yttrium aluminium garnet (YAG) laser, at a frequency of 2.5 Hz. The LSMO target was stoichiometric while for BFO, a target with nominal composition $Bi_{1.15}FeO_3$ was used, in order to compensate for the high volatility of Bi. (001)-oriented $SrTiO_3$ (STO) substrates were used. In the pseudo-cubic representation, BFO has a unit-cell parameter of 3.96 Å, LSMO of 3.88 Å and STO of 3.905 Å, so that the STO substrate induces a tensile strain on LSMO and a compressive strain on BFO. The optimal growth conditions for LSMO were previously determined to be of a deposition temperature $T_{dep}$=720°C and an oxygen pressure of $P_{dep}$=0.41 mbar [18], with a post-annealing at 300 mbar of $O_2$. We recently determined the $P_{dep}$-$T_{dep}$ phase stability diagram for BFO films and found that optimal conditions are around $T_{dep}$=580°C and $P_{dep}$=$6.10^{-3}$ mbar [19]. In the present samples, the LSMO layer (11 nm thick for all the samples) was grown first and the BFO film ($t_{BFO}$=1-70 nm) afterwards. To limit a possible deoxygenation of the manganite, after the growth of the LSMO film, the pressure was thus kept at the LSMO deposition pressure while decreasing $T_{dep}$ to 580°C. Then, the pressure was rapidly decreased to $6.10^{-3}$ mbar, to grow the BFO layer. The sample was finally cooled to room temperature in 300 mbar of oxygen.

Reflection High-Energy Electron Diffraction (RHEED) patterns were collected (at an acceleration voltage of 25 kV) before growth, and after depositing each layer. Images for the [100] direction are shown in figure 1a-c, and indicate a two-dimensional growth for the LSMO and the BFO layer (up to at least $t_{BFO}$=35 nm). An azimuthal analysis revealed an in-plane epitaxy for both layers.

X-ray diffraction θ-2θ spectra (in the 15°-115° 2θ range) were collected for all samples and only showed peaks corresponding to (00$l$) reflections (pseudo-cubic indexation) of STO, LSMO and BFO. Figure 1d shows the spectra for $t_{BFO}$=5 nm and 70 nm. From the angular position of the (003) reflections of BFO, we calculated the out-of-plane parameter $c_{BFO}$ ≈ 4.10 Å that was found to vary only slightly with thickness. A reciprocal space map (rsm) of the (103) reflections (see figure 1e) for a 70 nm film shows that both the LSMO and BFO layers are very heavily strained on the STO substrate. Note that no splitting of the BFO (103) peak is detected, suggesting a tetragonal rather than monoclinic or rhombohedral symmetry for the BFO layer, in contrast to the results of Xu et al [20] or Qi et al. [21] respectively.

Figure 2a shows a magnetization hysteresis cycle of a BFO(5nm)/LSMO bilayer measured at 10K with the field applied in-plane along [100] after zero field cooling, measured in a superconducting quantum interference device (SQUID). The saturation magnetization ($M_S$) for $t_{BFO}$=5nm is about 580 emu.cm$^{-3}$, close to the magnetization of bulk LSMO (590 emu.cm$^{-3}$). Even for larger BFO thickness values, the contribution of the BFO layer to the magnetization was not visible, as expected from the very small magnetization values obtained for optimized BFO single films grown in the same conditions [19]. We measured the evolution of the magnetization with the temperature in order to check the quality of the LSMO layer (see fig. 2b). The Curie temperature ($T_C$) is 330K, somehow smaller than the bulk $T_C$=370K [22], but in good agreement with the $T_C$ of thin single films [23] of similar thickness.

The BFO/LSMO bilayers were imaged with a conductive-tip atomic force microscope (CTAFM). In this type of experiments, the morphology of the sample surface and the resistance between the bottom electrode and the tip are measured simultaneously, and coupled height-resistance maps are recorded [24]. Examples of such maps (3x3μm²) are shown in figure 3a and 3b for a BFO(5nm)/LSMO bilayer. The left image reveals a very flat surface (in agreement with the two-dimensional growth mode inferred from the RHEED patterns) with terraces separated by ~ 4 Å-high steps, i.e. a perovskite unit-cell. The corresponding resistance map (figure 3b) shows a high resistance level (average value ~ 1 GΩ), with a remarkable homogeneity. Similar coupled maps were collected for samples with different BFO thicknesses. Identical unit-cell steps were observed up to $t_{BFO}$=20 nm. Resistance maps could be collected for the $t_{BFO}$=1, 2 and 5 nm samples, without saturating the capability of the CTAFM electronics. The average resistance of the maps is plotted as a function of the BFO thickness in figure 3c. An exponential increase of the resistance with $t_{BFO}$ is obtained, which indicates that transport occurs through the BFO film by tunnelling. This observation, together with the flatness and the homogeneity of the BFO on the LSMO buffers, qualifies BFO very thin films as potential barriers in tunnel junctions.

Piezoelectric atomic force microscopy (PFM) was used to probe the ferroelectricity of the BFO films in BFO/LSMO bilayers [25], for several values of $t_{BFO}$. An alternatively positive and negative voltage was applied between the conductive tip of the AFM and the bottom electrode (LSMO) to pole the BFO layer respectively "up" and "down". PFM was then used to detect the domain structure. Fig 4 shows the topography and PFM images of the 5 nm film after writing. We observe a clear contrast in the PFM image that reveals the presence of up and down ferroelectric domains in this film. Note that such pattern is not observed in the topography. These PFM measurements are not quantitative so that we cannot conclude on a possible decrease of the polarization of the BFO film when thickness decreases, as reported for $PbTiO_3$ for instance [26]. However, they clearly demonstrate that the ferroelectric character is preserved in BFO layers down to a thickness of 5 nm.

In summary, we have successfully grown epitaxial heterostructures integrating a LSMO bottom electrode and a BFO layer. The growth of both layers is two-dimensional, they are heavily strained and the surface of the structure is unit-cell flat. SQUID measurements indicate that the LSMO layer is ferromagnetic with a $T_C$ of 330K, while the BFO films give virtually no signal, as expected for a weak ferromagnet. Through a combined CTAFM and PFM study on these BFO/LSMO heterostructures, we have shown that BFO layers as thin as 5 nm are ferroelectric and can be used as tunnel barriers. This opens the way for the realisation of several types of devices, like ferroelectric tunnel junctions [15] or magnetic tunnel junctions [4] with ferroelectric tunnel barriers. In this latter type of structure, a control of the ferroelectric polarization by an external magnetic field can be envisaged, via the magnetoelectric effect.

We are very grateful to M. Viret and D. Colson for providing the BFO target and to J.-M. Triscone for fruitful discussions. G. Herranz and M. Sirena acknowledge financial support from the Ministère de l'Education Nationale, de l'Enseignement Supérieur et de la Recherche (France). H. Béa acknowledges financial support from the Conseil Général de l'Essone (France).

Figure Captions :

Fig 1. RHEED patterns in the [100] direction of the sample with $t_{BFO}$=20nm (a) of the STO substrate before deposition, (b) after deposition of the LSMO layer (c) after deposition of the BFO. The pattern is more diffuse in (b) because the pressure of measurement was 6 $10^{-3}$ mbar compared to $10^{-6}$ mbar for (a) and (c). (d) X-ray diffraction spectra of BFO/LSMO//STO bilayers with $t_{LSMO}$=11nm and $t_{BFO}$=5nm and 70nm. S, L, and B label peaks of STO, LSMO and BFO respectively. (e) Reciprocal space map of the (103) reflections of the BFO(70nm)/LSMO//STO bilayer; r.l.u. is for reciprocal space units.

Fig 2. (a) Magnetic hysteresis loops of the BFO/LSMO//STO bilayer with $t_{BFO}$=5nm measured by SQUID at 10K. (b) Temperature dependence of the magnetization in a field of 1kOe, normalized to the magnetization at 10K ($M_{10K}$) for the same sample.

Fig 3. 3x3µm² morphology (a) and resistance (b) maps of the BFO(5nm)/LSMO bilayer measured simultaneously by a conductive tip atomic force microscope. (c) Logarithm of the average resistance of these maps for different BFO thicknesses. The error bars correspond to the full width at half maximum of the resistance distribution.

Fig 4. Topography (a) and PFM image after the writing of domains using the AFM tip as a top electrode (b) of the 5nm BFO layer. The contrast in (b) indicates that the film is ferroelectric. The scale bar corresponds to 2 µm.

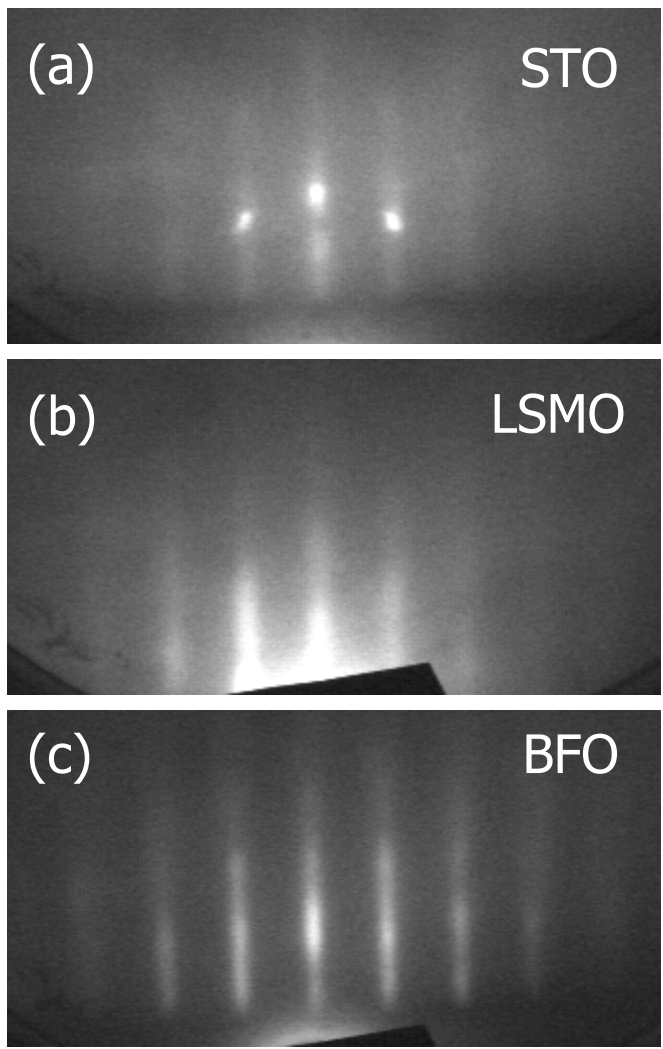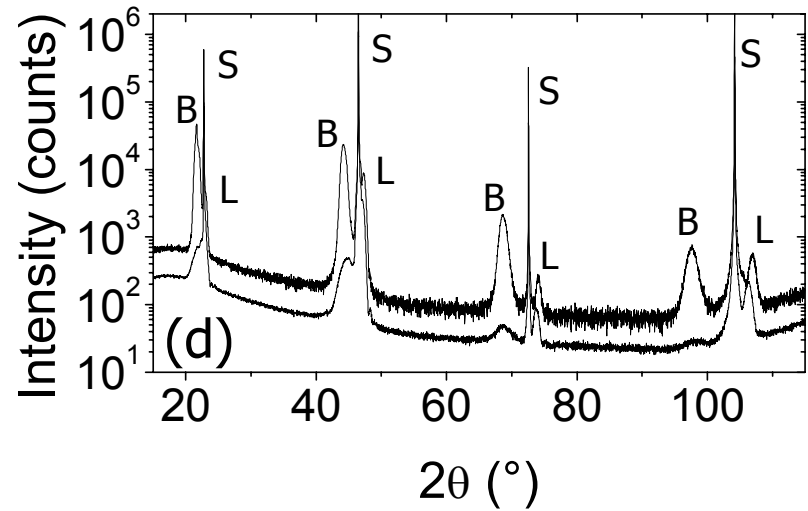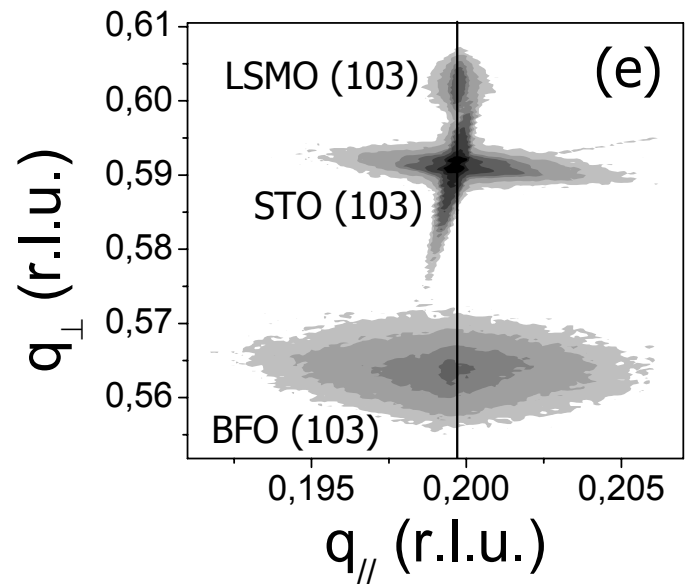

Béa et al.
Fig 1

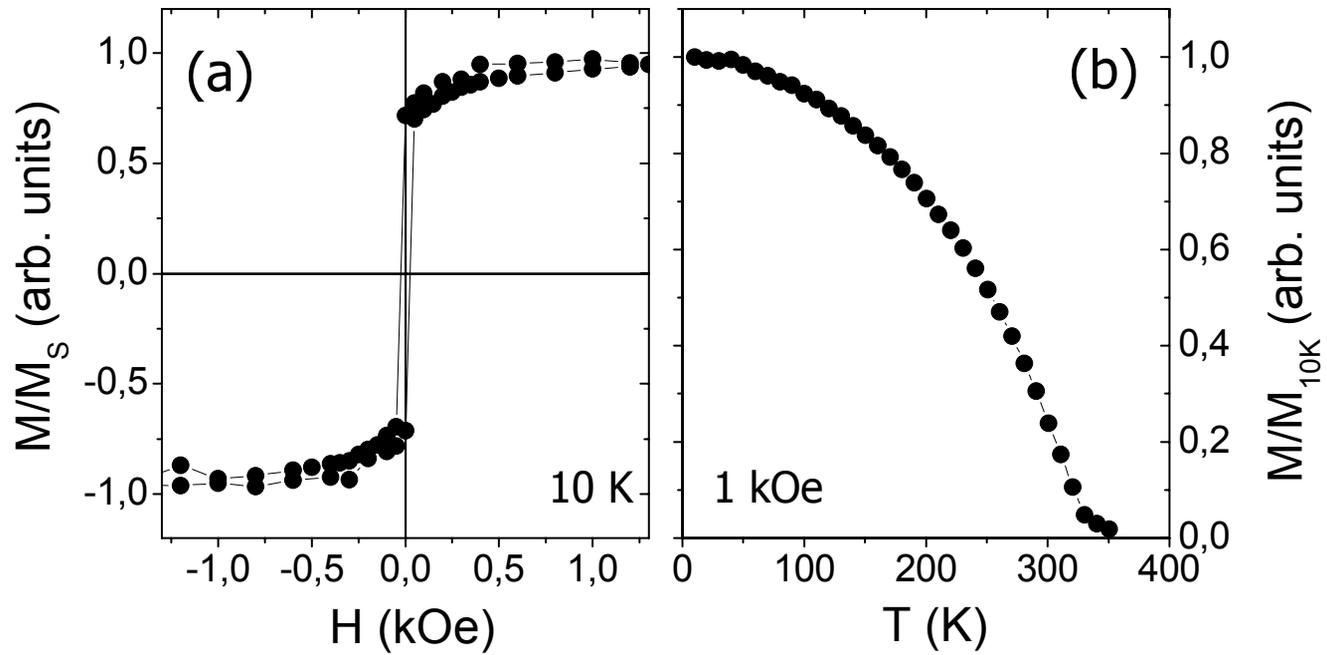

Béa et al.
Fig 2

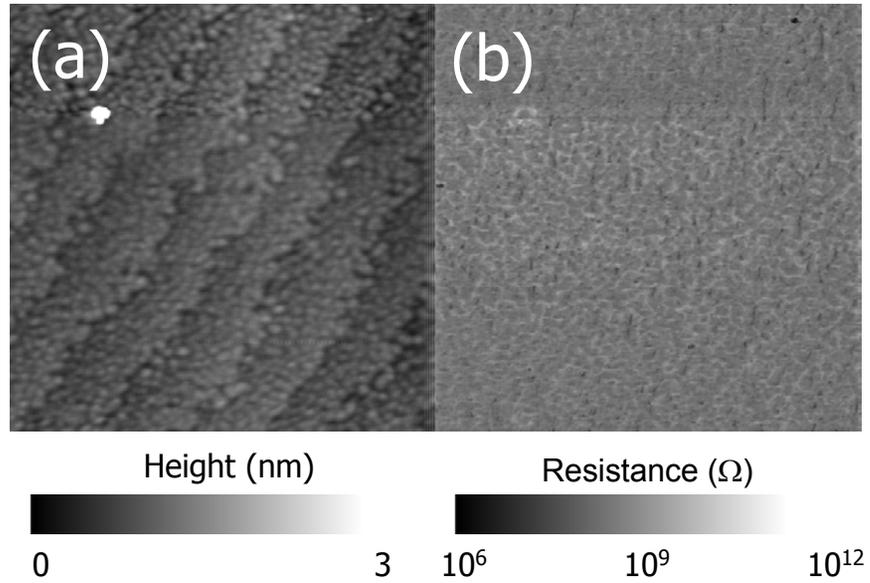
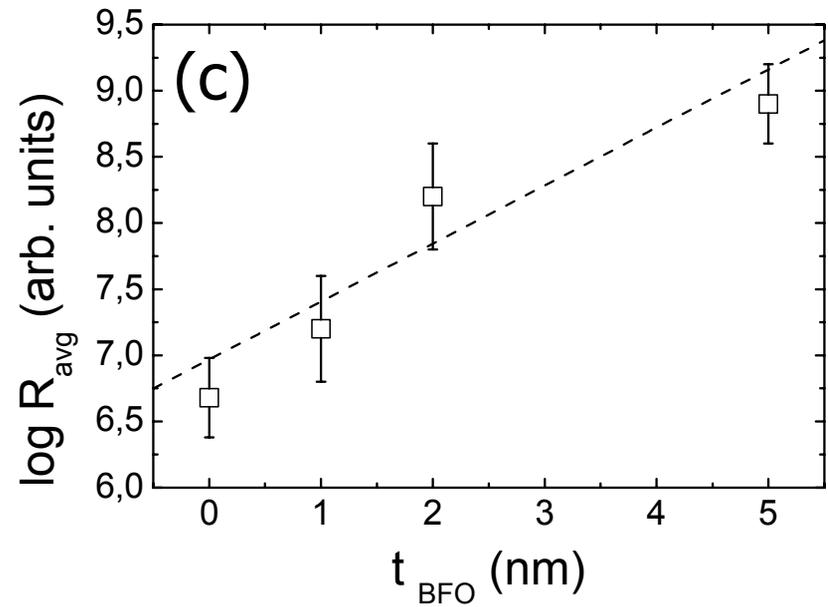

Béa et al.
Fig 3

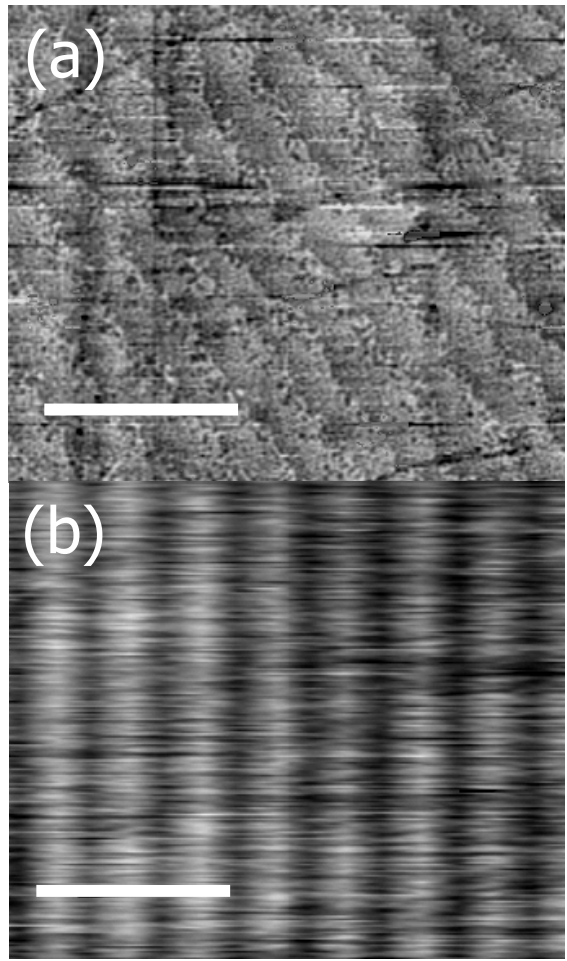

Béa et al.
Fig 4